\documentclass[twocolumn,prl,aps,superscriptaddress]{revtex4-1}
\usepackage[T1]{fontenc}
\usepackage[utf8]{inputenc}
\usepackage{xcolor}
\usepackage{pdfcolmk}
\usepackage{amsmath}
\usepackage{amssymb}
\usepackage{graphicx}
\PassOptionsToPackage{normalem}{ulem}
\usepackage{ulem}

\makeatletter

\providecolor{lyxadded}{rgb}{0,0,1}
\providecolor{lyxdeleted}{rgb}{1,0,0}

\DeclareRobustCommand{\lyxsout}[1]{\ifx\\#1\else\sout{#1}\fi}

\usepackage{bm}
\usepackage{color}

\newcommand{\beq}{\begin{equation}}
\newcommand{\eeq}{\end{equation}}
\newcommand{\bea}{\begin{eqnarray}}
\newcommand{\eea}{\end{eqnarray}}

\def\Im {\mbox{Im}}

\makeatother

\begin{document}
\title{Topological Graphene plasmons in a plasmonic realization of the Su-Schrieffer-Heeger Model }
\author{ Tatiana G. Rappoport}
\affiliation{Instituto de Telecomunicações, Instituto Superior Técnico, University
of Lisbon, Avenida Rovisco Pais 1, Lisboa, 1049001 Portugal}
\affiliation{Instituto de Física, Universidade Federal do Rio de Janeiro, Caixa
Postal 68528, 21941-972 Rio de Janeiro RJ, Brazil}
\author{ Yuliy V. Bludov}
\affiliation{Department and Centre of Physics, and QuantaLab, University of Minho,
Campus of Gualtar, 4710-057, Braga, Portugal}
\author{Frank H. L. Koppens}
\affiliation{ICFO-Institut de Ciencies Fotoniques, The Barcelona Institute of Science
and Technology, 08860 Castelldefels (Barcelona), Spain}
\affiliation{ICREA-Institució Catalana de Recerca i Estudis Avançats, Barcelona,
Spain}
\author{ Nuno M. R. Peres}
\affiliation{Department and Centre of Physics, and QuantaLab, University of Minho,
Campus of Gualtar, 4710-057, Braga, Portugal}
\affiliation{International Iberian Nanotechnology Laboratory (INL), Av. Mestre
José Veiga, 4715-330, Braga, Portugal}
\date{\today}
\begin{abstract}
Graphene hybrids, made of thin insulators, graphene, and metals can support propagating acoustic plasmons (AGPs). The metal screening modifies the dispersion relation of usual graphene plasmons leading to slowly propagating plasmons, with record confinement of electromagnetic radiation. Here, we show that a graphene monolayer, covered by a thin dielectric material and an array of metallic nanorods can be used as a robust platform to emulate the Su-Schrieffer-Heeger model. We calculate the Zak's phase of the different plasmonic bands to characterise their topology. The system shows bulk-edge correspondence: strongly localized interface states are generated in the domain walls separating arrays in different topological phases. We find signatures of the nontrivial phase which can directly be probed by far-field mid-IR radiation, hence allowing a direct experimental confirmation of graphene topological plasmons. The robust field enhancement, highly localized nature of the interface states, and their gate-tuned frequencies expand the capabilities of AGP-based devices.
\end{abstract}


\maketitle
Topology can lead to intriguing physical phenomena and it is at the
heart of modern condensed matter physics~\cite{RMPHasan,RMPZhang,revNiu,SZhang:2017}.
It has been successfully extended to various classical wave systems,
such as photonics~\cite{topphot1,RMPezawa,silveirinha15}, acoustic~\cite{topac}
and mechanical systems~\cite{Ma2019}. It also has been playing an
increasingly important role in nanophotononics~\cite{Rider2019},
offering alternative ways to design novel optical devices~\cite{Ota2020}.

In one dimension, the celebrated Su-Schrieffer-Heeger (SSH) is probably
the simplest and most representative model with non-trivial topology
\cite{ssh1,ssh2}. Originally, it describes electrons in a one-dimensional
tight-binding model with staggered hopping amplitudes, defined as
intracell and intercell hoppings \cite{ssh1}. Depending of the ratio
between the two hopping amplitudes, the chain can have two topologically
distinct ground states. The variation of this ratio leads to a topological
phase transition between the two phases, with the band gap closing
and reopening. If the intercell hopping is stronger than the intracell
hopping, the system is in a non-trivial topological phase. In this
case, the bulk-edge correspondence ~\cite{Rhim2017} predicts the
existence of end-states, and interface states when two lattices with
different topological phases are connected~\cite{RMPsolitons}.

Photonic and plasmonic systems provide a flexible platform for the
SSH model~\cite{Meng2014,Gao:15,Wang2016,henriques2020}. The effective
intracell and intercell hoppings can be controlled, for example, by
tuning distances via nanofabrication. Nontrivial topology in coupled
plasmonic nanoparticle arrays has been previously realized in 1D plasmonic
nanoparticle arrays~\cite{Ling2015,Downing2017,Gomez2017,Li2018,Pocock:2018}.
These systems, similar to the Su-Schrieffer-Heeger model, exhibit
highly localized edge states at their ends, which are robust against
perturbations~\cite{Pocock2019}. However, similarly to dielectric
photonic crystals, it is difficult to dynamically tune the 1D plasmonic
nanoparticle arrays and control their edge and interface states. To
overcome these limitations, one possibility is the use of highly tuneable
graphene plasmons~\cite{Jung2018,Fan2019}.

Graphene Plasmon-polaritons (GP) are vertically localized electromagnetic fields
(that is, surface waves) that can be excited
in both the mid-infrared (MIR) and
the Terahertz
(THz) spectral ranges.
They present oscillatory behavior at the interface
between graphene and a dielectric~\cite{Goncalves2016}. They can
exhibit high degree of spatial confinement when
compared to a wavelength of the same frequency in free space~\cite{Jablan2009,DeAbajo2014,Park2015,Koppens2014,Low2014,Goncalves2016,Celano2019}.
The configuration of Metal-Insulator-Graphene (MIG) for GPs~\cite{Alonso-Gonzalez2017,Lundeberg2017,Iranzo2018,Goncalves2016,Goncalves2020}
involving a thin insulating layer, can hold vertically confined modes with much larger momentum than normal graphene plasmons.
In this limit, the dispersion relation of the GP becomes
linear and the mode is known as acoustic graphene plasmon (AGP)\cite{Principi2011,Alonso-Gonzalez2017,Iranzo2018}.
With this hybrid system,
records in the spatial confinement of electromagnetic radiation has
been achieved ~\cite{Iranzo2018,Lee2019,Epstein2020}.

AGPs in periodic systems, e.g. involving periodic metallic rods on graphene (separated by an insulator), form plasmonic bands and present a wealth of different physical effects~\cite{Rappoport2020}. The AGP's lateral confinement that originates from the metallic nanostructure
mimics a plasmonic tight-binding model where graphene's gating and the
distance between rods can control the effective hopping, modifying
the band-structure and modulating the band gaps.
 
\begin{figure}[h]
\centering{\includegraphics[width=0.99\columnwidth]{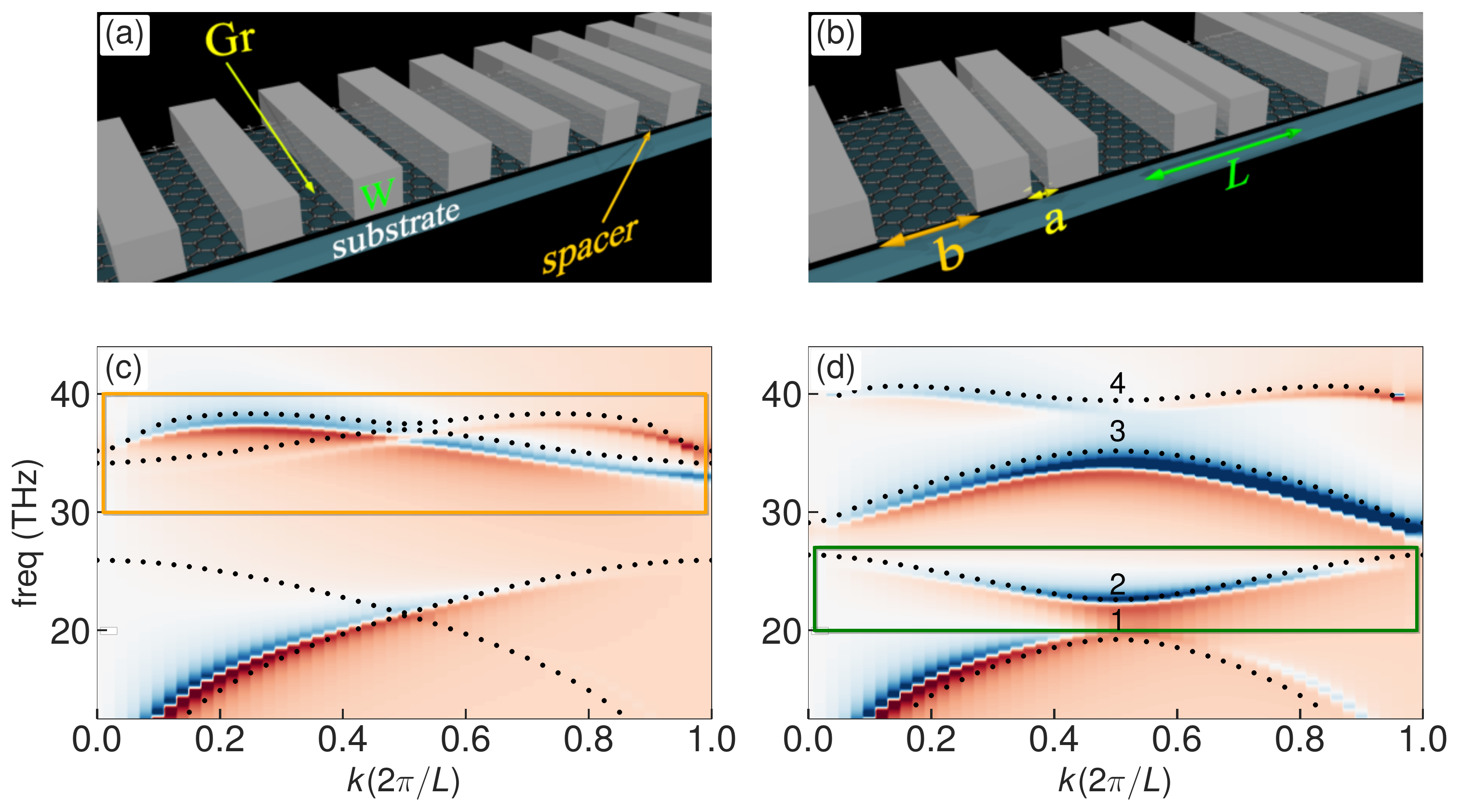}} \caption{\label{fig1} (a)-(b) Illustration of the one dimensional array of
rods with two rods per unit cell and $f=0$ and $f=2/3$ respectively. (c)-(d) Plasmonic band structure (dotted line) and loss function superimposed with the band structure (dotted lines), calculated for f=1/15 (c) and f=2/3 (d).  The yellow rectangle in (c)
highlights the band splitting at $k=\pi/L$ that occurs exactly at
the degenerate point in the band folding for $f=0$. The numbers in
panel (d) label the different bands}
\end{figure}

Here, we propose a novel one-dimensional topological graphene plasmonic
crystal that consists of a monolayer graphene on top of a bulk substrate
S with permittivity $\varepsilon_{S}$, separated from a periodic
structure of silver rods with cross-section of area 
$W^{2}=75\times75$ nm$^{2}$ by a thin dielectric spacer, of thickness $d=3$ nm and permittivity
$\varepsilon_{d}$ (see Fig \ref{fig1}a). The main advantage of
this structure is that   it is based on a recent experimental setup
to create graphene acoustic plasmons~\cite{Iranzo2018},
and therefore is experimentally feasible. The extra ingredient consists
in using two different separations between the rods, which can be
easily fabricated with the same techniques.  Furthermore, it avoids the use of metagates~\cite{Jung2018,Fan2019}.

As illustrated in Fig \ref{fig1}b, the 1D lattice unit cell contains
two identical silver rods separated by a distance $a$ and symmetrically
located with respect to the center of the unit cell. Neighboring rods
from different unit cells are separated by a distance $b$. The periodic
structure has a period $L=a+b+2W$. As $a$ dictates the intracell
effective hopping and $b$ is linked to the intercell effective
hopping, it is convenient to define the ratio $f=(a-b)/(a+b)$ that
controls the topology of our system. $f=0$ ( Fig. \ref{fig1}a)
implies the periodic structure studied previously~\cite{Rappoport2020}
with a single effective hopping. Positive (negative) values of $f$
specifies that the intracell effective hopping of our SSH model is
larger (smaller) than the intercell one, as shown in Fig \ref{fig1}b.

We perform full-wave finite element frequency domain simulations\cite{comsol}
and semi-analytical plane-wave expansions to characterize our
plasmonic SSH model (see Supplemetary Materials for the details~\cite{SM}). For simplicity, graphene is simulated as a single
layer with optical conductivity that is given by a Drude like expression $\sigma_{g}(\omega)=4\sigma_{0}E_{F}/(\pi(\hbar\gamma-i\hbar\omega))$~\cite{Goncalves2016},
where $\sigma_{0}=e^{2}/2\hbar$, $E_{F}$ is the Fermi energy, $\gamma$
is the relaxation rate and $\omega$ is the frequency of the incident
light. Since we will
be considering large graphene Fermi energies, finite temperature play
no significant effect in our results. The frequency-dependent relative
permittivities of Ag are taken from Ref.~\cite{Blaber2009}.

First, we analyse
the eigenfrequencies and eigenmodes of the system and distinguish
their topological phases for different values of $f$. We then proceed
to describe the interaction of EM radiation with our MIG structure.
We consider a $p$-polarized monochromatic plane-wave impinging on
the array of metallic rods at normal incidence. We calculate the absorption
spectra resulting from the coupling of AGPs with far-field radiation,
which can be used to design and model experiments. In this context,
we consider different setups: a periodic system, an interface of two
semi-infinite arrays with different topological phases, and 
edge states of a semi-infinite array interfacing a perfect electric
conductor (PEC) . Unless otherwise specified,
$\epsilon_{S}=\epsilon_{d}=1$, $E_{F}=0.6$ eV and $\gamma=3$ meV.

The dispersion of the plasmons in a periodic system with a single
rod per unit cell of length $L/2$ presents several plasmonic bands~\cite{Rappoport2020}.
If the same system is represented by a unit cell of length $L$ with
two evenly located rods ($f=0$), the dispersion can be depicted in
a Brillouin zone $k=[0,2\pi/L]$ which has half of the size of the
original one. As a result of the band folding, the dispersions cross
each other at $k=\pi/L$, which is a point of degeneracy. When calculating
the plasmonic band structure, from the analogy with a simple one-dimensional
tight-binding SSH model, we should expect a splitting of the original
bands for $f\neq0$ exactly at degenerate points of the band folding
(that is, at $k=\pi/L$),
with the size of the new gap being proportional to $|f|$. This can
be observed in our band structure and loss function calculations:
figure \ref{fig1}c shows a density plot of the loss function calculated
with the plane-waves expansion with a superimposed band-structure
for $f=1/15$. Each of the two original lowest bands for $f=0$ have
one degenerate point at $k=\pi/L$, which leads to a band folding
induced band gap, splitting the original bands in two. The yellow
rectangle in Fig. \ref{fig1}c highlights one of these band splittings,
which is very small, because of the value of $f$,
but illustrates nevertheless the effect of the band folding. The new Bragg gaps for $f\neq0$
are always located at $k=\pm\pi/L$. Fig. \ref{fig1}d presents the
same data but for $f=2/3$ and one can see that the band folding induced
gap increases for large values of $f$ and there are four well separated
bands labeled from 1 to 4. As expected, the band gap varies linearly
with $|f|$ for small values of $f$ (see the supplementary material~\cite{SM}), and the
band structures for $\pm f$ are exactly the same, although they correspond
to different topological phases.

The 1D array has chiral symmetry, as the unit cell consists of two interconnected sublattices (one for each rod)  that can be interchanged without modifying the system properties. One-dimensional periodical systems with chiral symmetry can be characterized
by a topological invariant known as Zak phase~\cite{Zak1989}. If
the unit cell has an inversion symmetry, the Zak phase is quantized
as $\pi$ (non-trivial) or 0 (trivial). To evaluate the Berry phase for electromagnetic waves in the absence of magneto-electric coupling, either the electric or magnetic fields can be considered in the calculation of the Berry connection $\vec{\Lambda}_{n,{\vec{k}}}$~ \cite{Onoda2006}. We adopted the electric field
in our calculations, where the permittivity tensor is isotropic and
given by $\hat{\epsilon}(\vec{r})=\epsilon(\vec{r})$. After considering
these simplifications, the Berry connection for an isolated band is
given by~\cite{Meng2014,Wang_2019}:

\begin{align}
{\vec{\Lambda}}_{n,{\vec{k}}}^{{\it {E}}} & ={\rm i}\int_{u.c}d{\vec{r}}\epsilon(\vec{r}){\vec{E}}_{n,{\vec{k}}}^{*}({\vec{r}})\cdot\nabla_{{\vec{k}}}{\vec{E}}_{n,{\vec{k}}}({\vec{r}}),
\end{align}
where $\vec{E}_{n,{\vec{k}}}({\vec{r}})$ is the periodic-in-cell
part of the normalized Bloch electric field eigenfunction of a state
on the $n$th band with wave-vector wavevector ${\vec{k}}$.

The periodic structure has periodicity in $\hat{x}$ and the system
is a one-dimensional plasmonic lattice. In this situation, the Zak
phase~\cite{Zak1989} is defined as $\theta_{n}=\int_{-\pi/L}^{\pi/L}dk\Lambda_{n,k}$.
The integral of the Berry connection over the BZ $-\pi/L\le k<\pi/L$
can be approximated as a summation of the contributions of small segments.
If the BZ is divided into $N$ segments where $k_{N+1}=k_{1}$, $e^{-{\rm i}\theta_{n}(k_{i})}\approx1-{\rm i}\theta_{n}(k_{i})=1-{\rm i}\Lambda_{n,k}\delta k$.
As we are dealing with plasmons confined in the region between the
metallic rods and the graphene sheet, without loss of generality,
we can calculate the Zak phase at a fixed height $z_{0}$ located
in the spacer with homogenous permittivity $\epsilon_{S}$. The Zak
phase of this segment $\theta_{n}(k_{i})$ for a band $n$ is given
by 
\begin{equation}
e^{-{\rm i}\theta_{n}(k_{i})}=\int_{u.c}d{x}{\vec{E}}_{n,{\vec{k}_{i}}}({x,z_{0}}){\vec{E}}_{n,{\vec{k}_{i+1}}}({x,z_{0}}),
\label{eq:theta}
\end{equation}

where $\theta_{n}$ can be calculated in a gauge-invariant formalism
as $\theta_{n}=-\Im[\log(\prod_{i=1}^{N}e^{-{\rm i}\theta_{n}(k_{i})})]$~\cite{Vanderbilt}.

\begin{figure}[h]
\centering{\includegraphics[width=0.99\columnwidth]{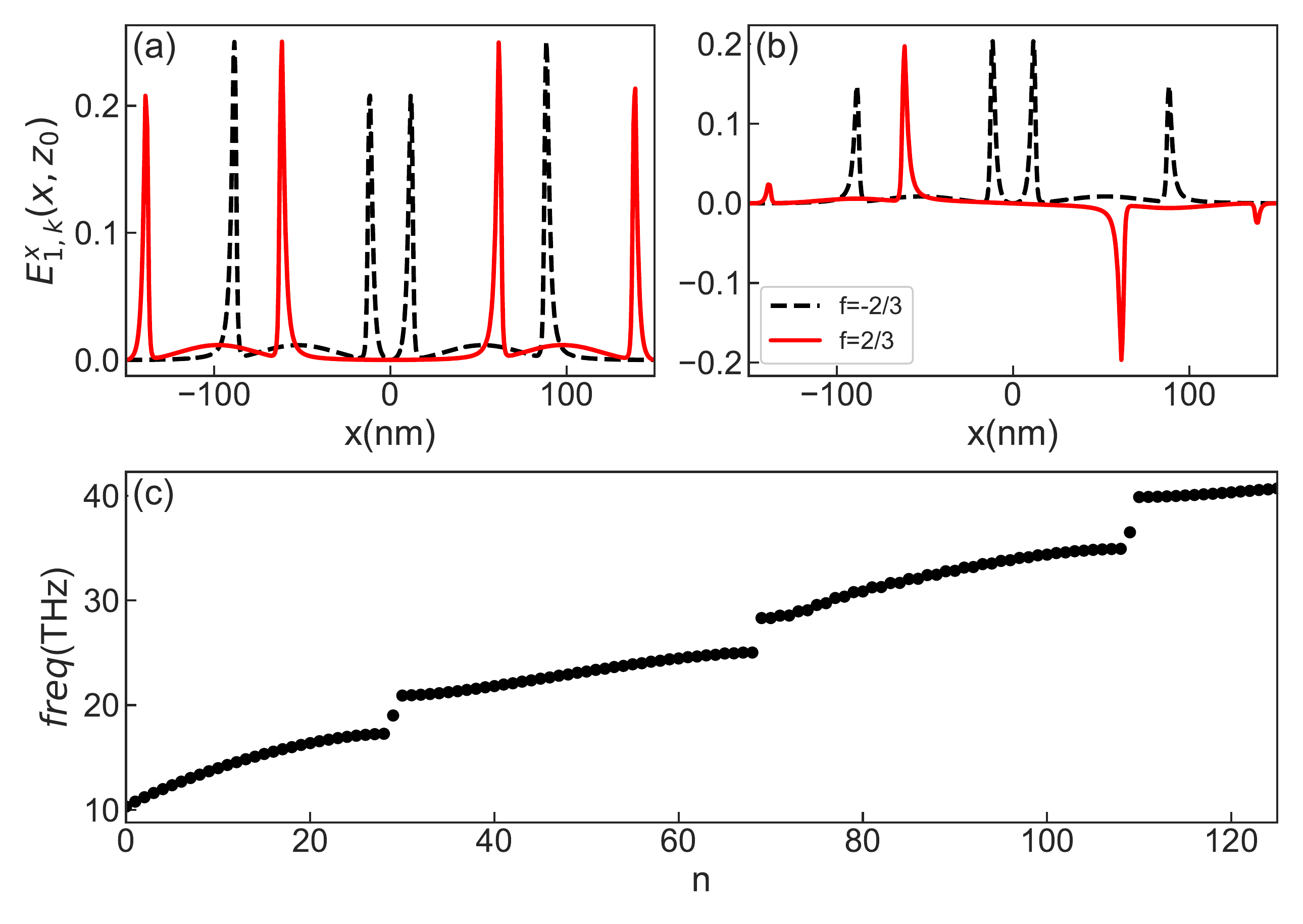}} \caption{\label{fig2} The periodic part of the longitudinal component of the
electric field $E_{n,{\vec{k}}}^{x}({x,z_{0}})$ for band 2, highlighted
in Figure \ref{fig1}d for $k=0$ (a) and $k=\pi/L$ (b) where $z_{0}$
is located in the middle of the spacer 2.0 nm above graphene and $-L/2\le x\le L/2$.
The dashed curves show the profiles for $f=-2/3$ while the profiles
of the solid curves are calculated for $f=2/3$. (c) Energy spectrum
of a composite system consisting of two
connected finite arrays of 20 unit cells each with $f=\pm2/3$ respectively, sandwiched by PECs. The mid-gap states are located in gaps after an
odd number of bands.}
\end{figure}

Alternatively, $\theta_{n}$ can be obtained by inspecting the parity
of the field profiles. If the symmetries of the eigenmode at $k=0$
and $k=\pm\pi/L$ are the same (different), the Zak phase of this
band is quantized as $0$ ($\pi$).  Following this procedure,
we illustrate the differences in the parity of the field profiles for band 2, highlighted
in green in Fig~\ref{fig1}d. Figure \ref{fig2}a, b presents
the field profile ${E^{x}}_{n,{k}}(x,z_{0})$  $k=0$ and $k=\pi/L$ respectively.  $z_{0}$ is located
in the middle of the spacer, between graphene and the rods. The phase
of the eigenmode is fixed in such a way that ${\vec{E}}_{n,{k=0}}(x,z_{0})$
is real. For $k=0$ (Fig. \ref{fig2}a) the field profiles for $f=\pm2/3$ are both even with respect to the inversion center of the unit cell, located at $x=0$. On the other hand,  for $k=\pi$ (Fig. \ref{fig2}b) the profile is even for $f=-2/3$ and odd for $f=2/3$. When comparing Fig. \ref{fig2}a and b it is clear that for $f<0$ ($f>0$) the symmetries of the eigenmodes for $k=0$ (a) and $k=\pi/L$(b) are  the same (opposite)  so that $\theta_n=0(\pi)$, which corresponds to the value obtained by
the Zak phase calculation following equation \ref{eq:theta}.

Lets us now address the physical consequences and experimental signatures
of the Zak phase in the graphene plasmonic crystal. To obtain a clear
signature of the topology, one route is the observation of interface
states for different values of $f$. Figure \ref{fig2}c shows the energy
spectrum for a single finite system consisting of two neighbouring arrays of 20 unit cells
each with $f=\pm2/3$ respectively. The system is sandwiched by perfect electric conductors (PECs). As discussed previously, systems with the same $|f|$ have the same spectrum. Consequently, both arrays have the same band-structure and the spectrum of the four lowest bands for the composite system is
similar to the unfolded version of the band structure of Fig \ref{fig1}d.
However, there is a clear presence of mid-gap states inside the Bragg
gaps. If two semi-infinite systems
with different topological phases form an interface, the existence
of a topological interface state in a given band gap is consistent
with the bulk-edge correspondence.   Thus, the  mid-gap states of Figure \ref{fig2}c are associated to the different Zak phases of each individual
array with $f=\pm2/3$, corroborating the previous analysis for periodic
systems. The original bands for $f=0$ are split in two for $f\neq0$ but they do not cross any other band, independent of the value of $f$.  Because of the conservation of topological numbers in band theory, the sum of the Zak phases of each pair of these bands is always
the same, regardless of the sign of $f$, although each individual band can change its phase when inverting the sign of $f$ and the Bragg gap goes to zero. This results in the absence of mid-gap states in gaps located after an even number of bands (see Fig \ref{fig2}c).

\begin{figure}[h]
\centering{\includegraphics[width=1\columnwidth]{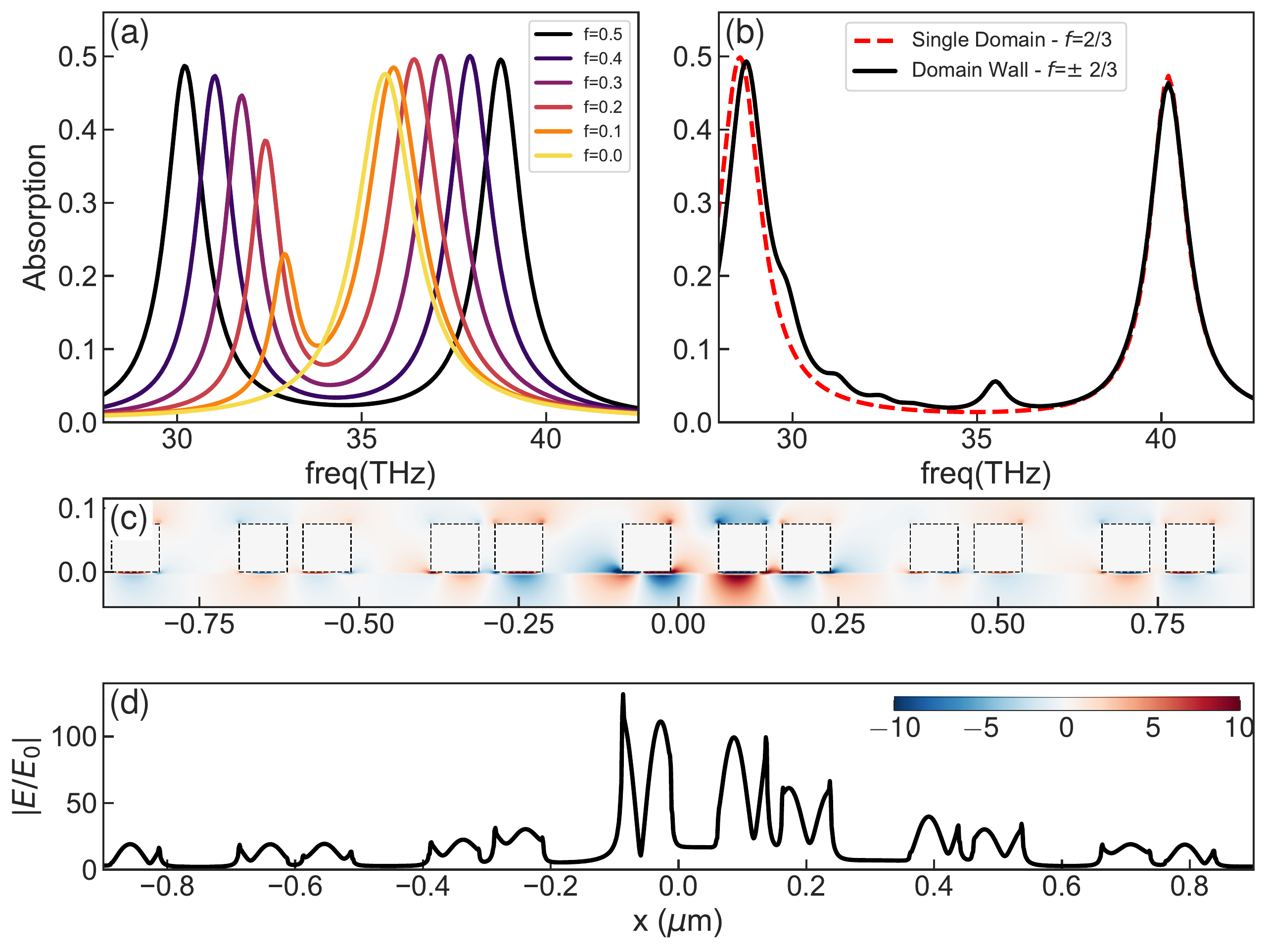}} \caption{\label{fig3} (a) Absorption spectra for a periodic array and different
values of $f=(a-b)/(a+b)$. (b) Absorption spectra of a finite plasmonic
crystal with $f=-2/3$  PEC boundary conditions (dashed lines) and the
interface containing two connected arrays of 6 unit cells each, with $f=\pm2/3$ (solid
line). (c) Electric field distribution $E^{y}(\vec{r})/E_{0}$ for
the region of the interface consisting of two connected arrays with
$f=\pm2/3$. The dashed squares denote the position of the metallic
rods. (d) Electric field enhancement $|\vec{E}(\vec{r})|/E_{0}$ in
the spacer between graphene and the rods. $x=0$ specifies the interface
between the two arrays with $f=\pm2/3$.}
\end{figure}

To explore the experimental signatures of the Zak phase, we consider
the coupling of the plasmonic crystal with far-field radiation. In
this case, we have a $p$-polarized monochromatic plane-wave impinging
on the array of metallic rods at normal incidence. Let us first consider
the periodic system and see if the splitting of the bands with different
values of $f$ can be observed in far-field experiments.  At normal incidence, TM modes
couple with states with $k_{x}\sim0$. Because of this, far-field
experiments cannot directly obtain the linear dependence of the gap
with $f$, as the gap opening associated to the SSH model occurs at
$k=\pi/L$. Still, it is possible to capture:
1) the existence of an extra band for $f\neq0$,
and 2) the band separation at $k_{x}=0$ for increasing values of
$|f|$. The first bands that can be seen in the far-field experiments
with normal incidence, are bands 3 and 4, highlighted in Fig \ref{fig1}
c. This is illustrated in figure \ref{fig3}a: the absorption spectra
for $f=0$ has a single peak at
this range of frequencies. For $f\neq0$, the band is split in bands
3 and 4 and produces two peaks in the absorption spectra, where their
position is dictated by the frequency of the bandstructure at $k=0$.
The peak separation is not directly related to the band gap but, instead,
to the values of the band structure for $k=0$. 

To analyse the interface states, we begin by considering a finite
plasmonic lattice with $f=-2/3$ with PEC ( perfect electric conductor) boundary conditions. This system presents exactly the same far-field response of
the periodic lattice (see Fig.~\ref{fig3}). We can now compare it
with the response for the interface considered in Figure \ref{fig2}c, involving two joined arrays with $f=\pm2/3$. In this case,
the interface state between different topological phases of the two
chain leads to an extra absorption peak located between the two original
peaks of the infinite system, seen in Fig. \ref{fig3}b. Figure
\ref{fig3}c shows the field enhancement at the interface state at
this particular frequency. The dashed squares indicate the position
of the metallic rods and the interface between the two different lattices
is located at $x=0$. One can see that in the vicinity of the domain
wall separating the two lattices, the field is enhanced in the whole
space, including the region above the rods. Figure \ref{fig3}d
exhibits the profile of $|\vec{E}(\vec{r})|/|\vec{E}_{0}|$, at $z_{0}$,
located in the spacer between graphene and the rods, where $\vec{E}_{0}$
is the  electrical field. In this case, the field enhancement
is of the order of 120-140, which is of the order of the field enhancement
of the main absorption peaks~\cite{Rappoport2020}.  The electric field profile in Fig. \ref{fig3}c 
has a maxima that is strongly localized in the region of the interface between the two arrays. This differs from
the extended electric field profile normally seen in the interface of photonic crystals 
with different Zak phases, where the interface state has a width of several unit cells \cite{Meng2018}.

The topological nature of this peak can be further confirmed by comparing
the absorption peaks of interfaces between two arrays with$f=f_{1}$
and$f=f_{2}$ where $|f_{1}|\neq|f_{2}|$. In this case, one can produce
interfaces between systems in the same topological phase (sgn($f_{1}$)=sgn($f_{2}$).
However, only interfaces between systems with sgn($f_{1}$)$\neq$
sgn($f_{2}$) produce interface states (see S. M). Although the topological
nature of the bands cannot be easily tuned in-situ, it is still possible
to use a gate to modify the optical conductivity in graphene. This
leads to a change in the size of the band widths and gaps and the
exact frequency of the edge state. The flexibility to modify the frequency
of the interface state can be useful for technological applications.

\begin{figure}[h]
\centering{\includegraphics[width=1\columnwidth]{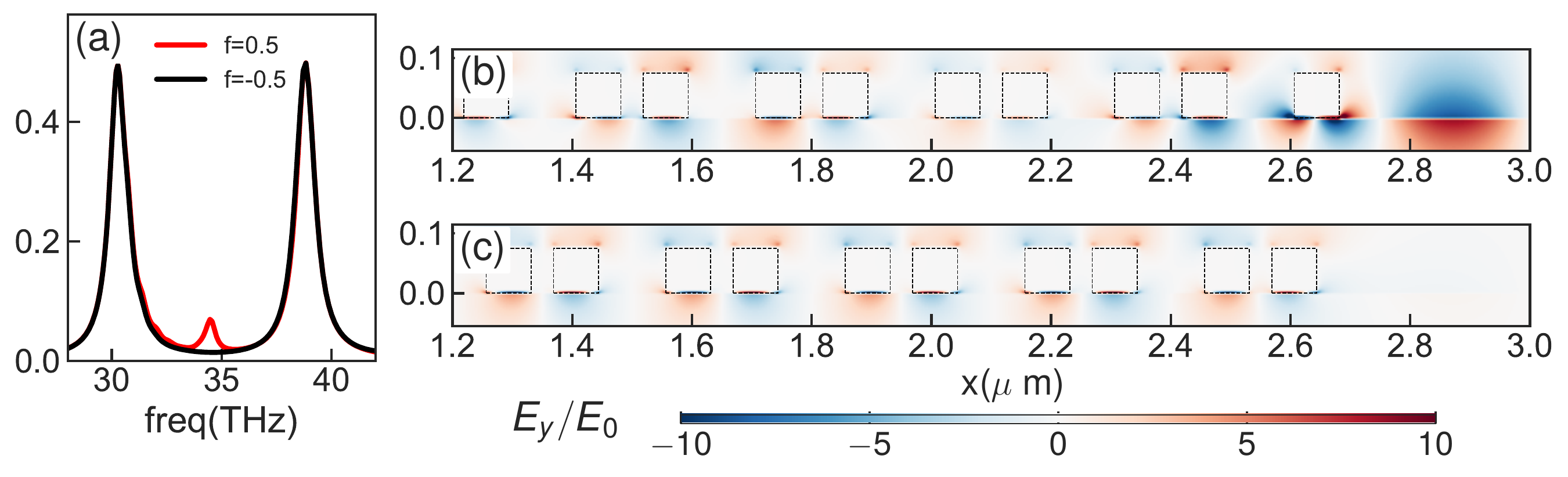}} \caption{\label{fig4} (a) Absorption spectra for a plasmonic crystal with
$f=\pm1/2$ with one vacancy interfacing a perfect electric conductor.
Electric field distribution $E^{y}(\vec{r})/E_{0}$ for lattices with
(b) $f=1/2$ and (c) $f=-1/2$ with one vacancy interfacing a perfect
electric conductor. The dashed squares denote the position of the
metallic rods. }
\end{figure}

We can now consider a vacancy or a void in the unit-cell neighboring
one of the PECs. This is obtained by removing the two silver rods
belonging to the last unit cell of the right, as illustrated in Fig.
\ref{fig4}b and c. In this case, the calculation of the eigenfrequencies
of the finite system do not produce in-gap states that are related
to the topology. Still, when analysing the interaction with far-field
radiation, different responses emerge, depending on the Zak phase
of the plasmonic band. For positive values of $f$, there is an extra
peak in the absorption spectra (Fig. \ref{fig4}a) which is related
to an interface state and electric field enhancement of the electric
field at the vacancy (Fig. \ref{fig4}b). However, for negative $f$s,
the absorption spectra is very similar to the periodic array and no
interface states are observed.  Instead, a resonator is formed between the last rod of the structure and the PEC. Depending on the size $D$ of the resonator, its eigenfrequency can be located inside the gap ($f>0$) or in the plasmonic band $f<0$), which explains the absorption spectra.

 Although we cannot connect this response
to topology, it can still be used in situations where one needs to
produce a field confinement in the absence of the metallic rod,
such as in sensing applications.

{ \textit{Conclusions} We proposed a simple structure to simulate
the Su-Schrieffer-Heeger Model for plasmons in graphene, which avoids the use of meta-gating . Our setup
is based on Metal-Insulator-Graphene systems that host acoustic graphene
plasmons. Periodic arrays of metallic rods with two rods per unit
cell generate plasmonic bands with topological properties that can
be tuned by the distances between the rods. Interface states with
strong field enhancement can be created at the interface between two
arrays with different topologies. The frequency of the localized state
can be easily tuned by gating graphene, which opens new avenues in
plasmonic applications where light needs to be confined to a precise
location in space or where a tunable narrow band absorption is needed.}

\begin{acknowledgements} TGR acknowledges funding from Fundação para
a Ciência e a Tecnologia and Instituto de Telecomunicações - grant
number UID/50008/2020 in the framework of the project Sym-Break and Mario G. Silveirinha for useful discussions.
 N.M.R.P. and F.H.L.K. acknowledge support from the European Commission through
the project “Graphene-Driven Revolutions in ICT and Beyond” (Ref.
No. 881603, CORE 3). N.M.R.P. acknowledge COMPETE 2020, PORTUGAL 2020,
FEDER and the Portuguese Foundation for Science and Technology (FCT)
through project POCI-01- 0145-FEDER-028114. F.H.L.K. acknowledges
financial support from the Government of Catalonia trough the SGR
grant, and from the Spanish Ministry of Economy and Competitiveness,
through the “Severo Ochoa” Programme for Centres of Excellence in
RD (SEV-2015- 0522), support by Fundacio Cellex Barcelona, Generalitat
de Catalunya through the CERCA program, and the Mineco grants Ramón
y Cajal (RYC-2012-12281, Plan Nacional (FIS2013-47161-P and FIS2014-59639-
JIN) and the Agency for Management of University and Research Grants
(AGAUR) 2017 SGR 1656. This work was supported by the ERC TOPONANOP
under grant agreement n 726001 and the MINECO Plan Nacional Grant
2D-NANOTOP under reference no FIS2016-81044-P. \end{acknowledgements}

%

 
\end{document}